\documentclass[preprint,aps,showpacs,amssymb,amsmath,superscriptaddress]{revtex4}
\usepackage{amssymb}
\usepackage[dvips]{graphicx}

\begin{document}
\title{High temperature superconductivity due to the long-range electron-phonon
interaction, application to isotope effects, thermomagnetic
transport and nanoscale heterogeneity in cuprates.}

\author{ A.S. Alexandrov}
 \affiliation{ Department of Physics, Loughborough University, Loughborough LE11\,3TU, UK\\
}

\begin{abstract}

Strong electron-phonon interactions in  cuprates and other
high-temperature superconductors have gathered support over the
last decade in a large number of experiments. Here  I briefly
introduce  the Fr\"ohlich-Coulomb multi-polaron model of
high-temperature superconductivity, which includes strong  on-site
repulsive correlations and the long-range Coulomb and
electron-phonon (e-ph) interactions.  Our extension of the BCS
theory to the strong-coupling regime with the long-range
\emph{unscreened} electron-phonon interaction  could naturally explain
 isotope effects,  unconventional  thermomagnetic transport,
and checkerboard modulations of the tunnelling density of states
in cuprates.
\end{abstract}

 \vskip2pc]

\maketitle

\section{Introduction: The Fr\"ohlich-Coulomb model}

A significant fraction of research in the field of
high-temperature superconductivity \cite{and2,kiv} suggests that
the interaction in novel superconductors is essentially repulsive
and unretarted, but it provides high $T_{c}$ without any phonons.
A motivation for this concept can be found in the earlier work by
Kohn and Luttinger \cite{koh}, who showed that the Cooper pairing
of repulsive fermions is possible. However, the same work showed
that $T_{c}$ of repulsive fermions is extremely low, well below
the mK scale. Nevertheless, the BCS and BCS-like theories
(including the Kohn-Luttinger consideration) heavily rely on the
Fermi-liquid model of the {\it normal }state. This model fails in
 cuprates, so that there are no obvious {\it a
priory} reasons to discard the dogma, if the normal state is not
the Fermi-liquid. Strong  onsite repulsive correlations (Hubbard
$U$) are essential in  undoped (parent)  cuprates, which
 are  {\it insulators }with the  optical gap about $2eV$ or so. Indeed,  if  repulsive
correlations are weak, one would expect a metallic behaviour of a
half-filled $d$-band of copper, or, at most, a much smaller gap
caused by lattice and spin distortions (i.e. charge and/or spin
density waves \cite{gab}). It is a strong onsite repulsion of
$d$-electrons  which results in the ``Mott'' insulating state of
parent cuprates. Different from conventional band-structure
insulators with completely filled or empty Bloch bands, the Mott
insulator arises from a potentially metallic half-filled band as a
result of  the Coulomb blockade of electron tunnelling to
neighboring sites \cite{mott}.

However, independent of any experimental evidence the Hubbard $U$
(or $t-J$)
 model
shares an inherent difficulty in determining the order. While some
groups claimed that it describes high-$T_{c}$ superconductivity at
finite doping, other authors could not find any superconducting
instability without an additional (i.e. e-ph) interaction
\cite{scher}. Therefore it has been concluded that models of this
kind are highly conflicting and confuse the issue by exaggerating
the magnetism rather than clarifying it \cite{lau}. There is
another serious problem with the Hubbard-U approach to high
temperature superconductivity in  cuprates.  The characteristic
(magnetic) interaction, which \emph{might} be responsible for the
pairing in the Hubbard  model, is the spin exchange interaction,
$J=4t^{2}/U$, of the order of $0.1$ eV.  It turns out much smaller
than the
 (inter-site) Coulomb repulsion and the unscreened long-range
(Fr\"ohlich) e-ph interaction each of the order of 1 eV, routinely
neglected within the approach. There is virtually no screening of
 e-ph interactions with $c-$axis polarized optical phonons in
cuprates because the upper limit for the out-of-plane plasmon
frequency ($\lesssim 200$ cm$^{-1}$)\cite{bas} is well below the
characteristic phonon frequency, $\omega\approx$ 400 - 1000 cm
$^{-1}$ . As a result of  poor screening, the magnetic interaction
remains small compared with the Fr\"ohlich interaction at any
doping. Hence, any realistic approach to superconductivity and
heterogeneity in cuprates should treat the Coulomb and
\emph{unscreened} e-ph interactions on an equal footing.

We have developed a so-called ``Fr\"ohlich-Coulomb'' model
\cite{alebook,alekor} to deal with the strong Coulomb and
long-range e-ph interactions in cuprates and other doped oxides.
The model Hamiltonian explicitly includes the long-range
electron-phonon and Coulomb interactions as well as the kinetic
and deformation energies.  The implicitly present large Hubbard
term prohibits double occupancy and removes the need to
distinguish fermionic spins. Introducing spinless fermionic,
$c_{\bf n}$, and phononic, $d_{{\bf m}\alpha }$, operators the
Hamiltonian of the model  is written as
\begin{eqnarray}
\tilde{H} = & - & \sum_{\bf n \neq n'} \left[ t({\bf n-n'}) c_{\bf
n}^{\dagger } c_{\bf n'} - V_{c}({\bf n-n'}) c_{\bf n}^{\dagger}
c_{\bf n}c_{\bf n'}^{\dagger } c_{\bf n'} \right]  \nonumber \\
& - &  \sum_{\bf n m} \omega_{\alpha} g_{\alpha}({\bf m-n}) ({\bf
e}_{\alpha } \cdot {\bf u}_{\bf m-n}) c_{\bf n}^{\dagger } c_{\bf
n} (d_{{\bf m}\alpha}^{\dagger}+d_{{\bf m}\alpha }) +
 \sum_{{\bf m}\alpha} \omega_{\alpha}\left( d_{{\bf m}\alpha
}^{\dagger} d_{{\bf m}\alpha }+\frac{1}{2} \right),
\end{eqnarray}
where ${\bf e}_{ \alpha}$ is the polarization vector of $\alpha$th
vibration coordinate, ${\bf u}_{\bf m-n} \equiv ({\bf m-n})/|{\bf
m-n}|$ is the unit vector in the direction from electron ${\bf n}$
to the ion ${\bf m}$, $g_{\alpha}({\bf m-n)}$ is a dimensionless
e-ph coupling function, and $V_{c}({\bf n-n'})$ is the inter-site
Coulomb repulsion. $g_{\alpha}({\bf m-n)}$ is proportional to the
{\em force} acting between an electron on site ${\bf n}$ and an
ion ${\bf m}$. For simplicity, we assume that all the phonon modes
are non-dispersive with the frequency $\omega_{\alpha}$. We also use
$\hbar = 1$.

The  Hamiltonian, Eq.(1), can be solved analytically in the
extreme case of a strong e-ph interaction where the e-ph
dimensionless coupling constant is large, $\lambda =E_p/zt >1$,  by using
$1/\lambda$ multipolaron expansion technique \cite{alebook}. Here
$ E_{p} =  \sum_{{\bf n} \alpha} \omega_{\alpha}
g_{\alpha}^{2}({\bf n}) ({\bf e}_{\alpha}\cdot {\bf u}_{\bf
n})^{2} ,$ is the polaron level shift about 1 eV and $zt$ is the
half-bandwidth in a rigid lattice.

The model shows a reach phase diagram depending on the ratio of
the inter-site Coulomb repulsion $V_{c}$ and the polaron
(Franc-Condon) level shift $E_{p}$ \cite{alekor}. The ground state
is a \emph{polaronic} Fermi  liquid at large Coulomb repulsions, a
\emph{bipolaronic} high-temperature superconductor at intermediate
Coulomb repulsions, and a charge-segregated insulator at weak
repulsion.  The model predicts \emph{superlight } bipolarons with
a remarkably  high superconducting critical temperature. It
describes many other properties of 
 cuprates \cite{alebook}, in particular band-structure isotope
effects, normal state thermomagnetic transport and real-space
modulations of the single-particle density of states (DOS) as
discussed below.

 \section{Band-structure isotope effect}

The isotope substitution, where an ion mass $M$ is varied without
any change of electronic configurations, is a powerful tool in
testing the origin of electron correlations in solids. In
particular, a finite value of the isotope exponent $\alpha =-d\ln
T_{c}/d\ln M$  proved that the superconducting phase transition at
$T=T_c$ is driven by the electron-phonon interaction in
conventional low-temperature superconductors \cite{iso}. Advances
in  fabrication of isotope substituted samples made it possible
to measure a sizable isotope effect also in many high-temperature
superconductors. This led to a general conclusion that phonons are
relevant for high $T_{c}$. Essential features of the isotope
effect on $T_c$, in particular its large values in low $T_{c}$
cuprates, an overall trend to decrease as $T_{c}$ increases, and a
small or even negative $\alpha $ in some high $T_{c}$ cuprates,
could be understood in the framework of the bipolaron theory of
high-temperature superconductivity \cite{aleiso}.

The most compelling evidence for (bi)polaronic carries in novel
superconductors was provided by the discovery of a substantial
isotope effect on the (super)carrier mass \cite{guo0,guo,kha}. The
effect was observed by measuring the magnetic field penetration
depth $\lambda_{H}$ of isotope-substituted copper oxides. The
carrier density is unchanged with the
isotope substitution of O$^{16}$ by O$^{18}$, so that the isotope effect on $%
\lambda_{H}$ measures directly the isotope effect on the carrier mass $%
m^{\ast }$. A carrier mass isotope exponent $\alpha_{m}=d\ln
m^{\ast }/d\ln M $ was observed, as predicted by the bipolaron
theory \cite{aleiso}. In ordinary metals, where the Migdal
adiabatic approximation \cite{mig} is believed to be valid,
$\alpha _{m}=0$ is expected. However, when the e-ph interaction is
sufficiently strong and electrons form polarons, their effective
mass $m^{*}$ depends on $M$ as $m^{*}= m \exp (\gamma
E_p/\omega)$. Here $m$ is the band mass in the absence of the
electron-phonon interaction, $\gamma$ is a numerical constant less
than $1$ that depends on the radius of the electron-phonon
interaction, and $\omega$ is a
characteristic phonon frequency. In the expression for $%
m^*$, only the phonon frequency depends on the ion mass. Thus
there is a
large isotope effect on the carrier mass in (bi)polaronic conductors, $%
\alpha_{m} = (1/2)\ln (m^*/m)$ \cite{aleiso}, in contrast with the
zero isotope effect in ordinary metals.

Recent high resolution angle resolved photoemission spectroscopy
(ARPES) \cite{lan0} provided another compelling evidence for a
strong e-ph interaction in  cuprates. It revealed a fine phonon
structure in the electron self-energy of the underdoped
La$_{2-x}$Sr$_x$CuO$_4$ samples \cite {shencon}, and a complicated
isotope effect on the electron spectral function in Bi2212
\cite{lan}. Polaronic carriers were also observed in colossal
magneto-resistance manganite including their low-temperature
ferromagnetic phase, where the isotope effect on the residual
resistivity was measured and explained \cite{alezhao}.

These and many other experimental and theoretical observations
point towards unusual e-ph interactions in complex oxides, which
remain to be quantitatively addressed. We have performed quantum
Monte Carlo (QMC) simulations \cite{alekor1} of a single-polaron
problem with the long-range e-ph interaction, Eq.(1), in the most
relevant intermediate region of the coupling strength, $\lambda
\simeq 1$, and of the adiabatic ratio, $\omega/t \simeq 1$, where
any analytical or even semi-analytical approximation (i.e. dynamic
mean-field approach in finite dimensions) may fail. More
recently we have calculated the isotope effect on the whole
polaron band dispersion, $\epsilon_{\mathbf{k}}$, and DOS of a
two-dimensional lattice polaron with short- and long-range e-ph
interactions by applying the continuous-time QMC algorithm
\cite{alekor2}. Unlike the strong-coupling limit \cite{aleiso} the
isotope effect depends on the wave vector in the intermediate
region of parameters. It also depends on the radius of the e-ph
interaction. If we define a band-structure isotope exponent
$\alpha_b$ as
\begin{equation}
\alpha_b \equiv - \frac{\partial \ln
\epsilon_{\mathbf{k}}}{\partial \ln M},
\end{equation}
it does not depend on the wave-vector $\mathbf{k}$ in the extreme
strong-coupling limit, $\lambda \gg 1$. E-ph interactions
do not change the band topology in this limit \cite{aleiso}, so
that $\alpha_b$ is the same as $\alpha_m$. Notwithstanding QMC
results \cite{alekor2}  show that the isotope exponent becomes a
nontrivial function of the wave vector, $\alpha _{b} = \alpha
_{b}(\mathbf{k)}$, in the intermediate-coupling regime, because
e-ph interactions substantially modify the band topology in this
 regime.

The isotope exponent $\alpha _{b}(\mathbf{k})$  are presented in
Figs.~(1,2),  for the small Holstein polaron (SHP) with the
short-range interaction, and for the small Fr\"ohlich polaron
(SFP) with the long-range interaction \cite{alekor2}. The polaron
spectra are calculated for two phonon frequencies, $\omega
=0.70\,t$ and $\omega
=0.66\,t $, whose difference corresponds to a substitution of O$^{16}$ by O$%
^{18}$ in cuprates. There is a significant change in the
dispersion law (topology) of SHP, Fig.1, which is less significant
for SFP, Fig.2, rather than a simple band-narrowing.  As a result,
the isotope exponent, $\alpha _{b}(\mathbf{k})\approx 8[\epsilon _{\mathbf{k}}^{16}-\epsilon _{%
\mathbf{k}}^{18}]/\epsilon _{\mathbf{k}}^{16}$,
is $\mathbf{k}$-dependent, Fig.1 and Fig.2 (lower panels). The strongest dispersion of $%
\alpha _{b}$ is observed for SHP. Importantly, the isotope effect
is suppressed near the band edge in contrast with the
$\mathbf{k}$-independent strong-coupling isotope effect. It is
less dispersive for SFP, especially at larger $\lambda$.

The coherent motion of small polarons leads to metallic conduction
at low temperatures. Our results, Figs.1,2  show that there should
be anomalous isotope effects on low-frequency kinetics and
thermodynamics of polaronic conductors which  depend on the
position of the Fermi level in the polaron band. Actually, such
effects have been observed in ferromagnetic oxides at low
temperatures \cite{alezhao}, and in cuprates \cite{guo0,guo,kha}. To
address ARPES isotope exponents \cite{lan} one has to calculate
the electron spectral function $A(\mathbf{k}, E)$ taking into
account phonon side-bands (i.e. off diagonal transitions) along
with the coherent polaron motion (diagonal transitions). Using the
$1/\lambda$ expansion and a single phonon mode approximation one
obtains \cite{alebook}
\begin{equation}
A(\mathbf{k},E )=\sum_{l=0}^{\infty } \left[ A_{l}^{(-)}(\mathbf{k}%
,E)+A_{l}^{(+)}(\mathbf{k},E )\right] ,  \label{eq:eight}
\end{equation}
where
\begin{eqnarray}
&& A_{l}^{(-)}(\mathbf{k},E ) = \frac{Z\left[ 1-n(E-l\omega) \right]}{%
(2N)^{l}l!} \cr &\times& \sum_{\mathbf{q}_{1},...\mathbf{q}%
_{l}}\prod_{r=1}^{l}|\gamma (\mathbf{q}_{r})|^{2} \delta [ E -
l\omega - \xi (\mathbf{k}_l^-)] ,  \nonumber  \label{eq:nine}
\end{eqnarray}
and
\begin{equation}
A_{l}^{(+)}(\mathbf{k},E ) = \frac{Z \cdot
n(E+l\omega)}{(2N)^{l}l!}
\sum_{\mathbf{q}_{1},...\mathbf{q}_{l}} \prod_{r=1}^{l}| \gamma (\mathbf{q}%
_{r})|^{2} \delta [ E - l\omega - \xi (\mathbf{k}_l^+)].
\label{eq:ten}
\end{equation}
Here $Z=\exp(-E_p/\omega)$, $\xi(\mathbf{k})=\epsilon_{\mathbf{k}}-\mu$, $%
\mu$ is the chemical potential, $n(E)=[\exp (\beta E)+1]^{-1}$, $\mathbf{k}%
_l^\pm=\mathbf{k}\pm \sum_{r=1}^{l} \mathbf{q}_r$ and
$\gamma(\mathbf{q})$ is the Fourier transform of the force,
$\gamma(\mathbf{q})=\sqrt{2}\sum_{\mathbf{n}}g({\bf n}) ({\bf e}
\cdot {\bf u}_{\bf n}) \exp(i\mathbf{q\cdot n})$.

Clearly, Eq.~(\ref{eq:eight}) is in the form of a perturbation
multi-phonon expansion. Each contribution $A_{l}^{(\pm
)}(\mathbf{k},E )$ to the spectral function describes the
transition from the initial state $\mathbf{k}$ of the polaron band
to the final state $\mathbf{k}_l^\mp$ with the emission (or
absorption) of $l$ phonons. Different from the conventional Migdal
self-energy \cite{mig} the electron spectral function comprises
two  parts in the strong-coupling limit. The first
($l=0$) $\mathbf{k}$-dependent term arises from the coherent
polaron tunnelling, $A_{coh}(\mathbf{k},E )= \left[
A_{0}^{(-)}(\mathbf{k},E ) + A_{0}^{(+)}(\mathbf{k},E ) \right] =
 Z \delta(E -\zeta_{\mathbf{k}})$ with a suppressed spectral
weight $Z \ll 1$. The second \textit{incoherent} part
$A_{incoh}(\mathbf{k},E )$ comprises all terms with $l \geq 1$. It
describes excitations accompanied by emission and absorption of
phonons. We note that its spectral density spreads over a wide
energy range of about twice the polaron level shift $E_{p}$, which
might be larger than the unrenormalized bandwidth $2zt$ in the
rigid lattice. On the contrary, the coherent part shows a
dispersion only in the energy window of the order of the polaron
bandwidth. It is important that the $incoherent$
background $A_{incoh}(\mathbf{k},E )$ could be dispersive (i.e. $\mathbf{k}$%
-dependent) for the long-range interaction. Only in the Holstein
model with the short-range nondispersive e-ph interaction,
$\gamma(\mathbf{q}) = constant$, the incoherent part has no
dispersion.

Using Eq.~(\ref{eq:eight}) one readily predicts the isotope effect
on the coherent part dispersion $\epsilon_{\mathbf{k}}$ and its
spectral weight $Z$, \emph{and}
also on the incoherent background because $Z$, $\gamma(\mathbf{q})$, and $%
\omega$ all depend on $M$. While our prediction is qualitatively
robust it is difficult to quantify the ARPES isotope effect in the
intermediate region of parameters. The spectral function,
Eq.~(\ref{eq:eight}), is applied in
the strong-coupling limit $\lambda \gg1$. While the main sum rule $%
\int_{-\infty }^{\infty }dE A(\mathbf{k},E )=1$ is satisfied, the
higher-momentum integrals $\int_{-\infty }^{\infty }dE E^{p}
A(\mathbf{k},E ) $ with $p>0$, calculated with
Eq.~(\ref{eq:eight}) differ from their exact values \cite{kor2} by
an amount proportional to $1/\lambda$. The difference is due to
partial ``undressing'' of high-energy excitations in phonon
side-bands, which is beyond the lowest order $1/\lambda$
expansion. The role of electronic correlations should be also
addressed in connection with ARPES. While the results shown in
Figs.1,2 describe band-structure isotope effects in slightly-doped
conventional and Mott-Hubbard insulators with a few carriers,
their spectral properties could be significantly modified by the
polaron-polaron interactions \cite{dev}, including the bipolaron
formation \cite{aledent} at finite doping. On the experimental
side, separation of the coherent and incoherent parts in ARPES
remains rather controversial.

\section{Thermomagnetic transport}

There is much evidence for the crossover regime at $T^{\ast }$ and
normal state charge and spin gaps in the cuprates
\cite{alebook,tim}. Within the Fr\"ohlich-Coulomb model these
energy gaps could be understood as being half of the binding
energy, $\Delta_{p},$ and the singlet-triplet gap of preformed
bipolarons, respectively \cite{alebook}. Notwithstanding some
"direct" evidence for the existence of a charge $2e$ Bose liquid
in the normal state of cuprates is highly desirable.  Mott and
Alexandrov\cite {NEV} discussed the thermal conductivity $\kappa
$; the contribution from the carriers given by the Wiedemann-Franz
ratio depends strongly on the elementary charge as $\sim (e^{\ast
})^{-2}$ and should be significantly suppressed in the case of
$e^{\ast }=2e$ compared with the Fermi-liquid contribution. As a
result, the Lorenz number, $L=\left( e/k_{B}\right)
^{2}\kappa _{e}/(T\sigma )$ differs significantly from the Sommerfeld value $%
L_{e}=\pi ^{2}/3$ of the standard Fermi-liquid theory, if carriers
are bipolarons. Here $\kappa _{e}$, $\sigma $, and $e$ are the
electronic thermal conductivity, the electrical conductivity, and
the elementary
charge, respectively. Ref. \cite{NEV} predicted a rather low Lorenz number $%
L_{b}$ for bipolarons, $L_{b}=6L_{e}/(4\pi ^{2})\approx
0.15L_{e}$, due to the double charge on bosonic carriers, and also due to
their nearly classical distribution  above $T_{c}$.

Unfortunately, the extraction of the electron thermal conductivity
has proven difficult since both the electron term, $\kappa _{e}$
and the phonon term, $\kappa _{ph}$ are comparable to each other
in the cuprates. Only recently a new way to determine the Lorenz
number has been realized by Zhang et al.\cite{zha}, based on the
thermal Hall conductivity. The thermal Hall effect allowed for an
efficient way to separate the phonon heat current
even when it is dominant. As a result, the ``Hall'' Lorenz number, $%
L_{H}=\left( e/k_{B}\right) ^{2}\kappa _{xy}/(T\sigma _{xy})$, has
been
directly measured in $YBa_{2}Cu_{3}O_{6.95}$ because transverse thermal $%
\kappa _{xy}$ and electrical $\sigma _{xy}$ conductivities involve
only  electrons. Remarkably, the measured value of $L_{xy}$
just above $T_{c}$ is about the same as predicted by the bipolaron
model, $L_{xy}\approx 0.15L_{e}. $ 

The experimental $L_{xy}$
showed a strong temperature dependence, which violates the
Wiedemann-Franz law. This experimental observation is hard to
explain in the framework of any Fermi-liquid model. Based on the Fr\"ohlich-Coulomb model and the conventional
Boltzmann kinetics in the $\tau$ approximation we developed a
theory of the Lorenz number in the cuprates explaining the
experimental results by Zhang et al \cite{lor}. We have
demonstrated that the Wiedemann-Franz law breaks down because of
the interference of polaron and bipolaron contributions to the
heat transport. When thermally excited polarons and also triplet
pairs are included, our model explains the violation of the
Wiedemann-Franz law in  cuprates and the Hall Lorenz number as
seen in the experiment, Fig.3.

A large Nernst signal observed \emph{well above} the resistive
critical temperature  $T_{c}$ provides another important piece of
evidence for a qualitatively different normal state of cuprates as
compared with conventional superconductors. It has been attributed
to a \emph{vortex} motion in a number of cuprates \cite{xu,wang1}.
As a result the magnetic phase diagram of the cuprates has been
revised with the upper critical field $H_{c2}(T)$ curve not ending
at $T_{c0}$ but at much higher temperatures  \cite{wang1}. Most
surprisingly, Ref.\cite{wang1} estimated $H_{c2}$ \emph {at the
zero-field resistive transition temperature} of Bi2212, $T_{c0}$,
as high as 50-150\,Tesla. However, any scenario with a nonzero
off-diagonal order parameter (the Bogoliubov-Cor'kov anomalous
average), $F({\bf r}_{1},{\bf r}_{2})= \left\langle \hat {\Psi}
_{\downarrow}({\bf r}_{1})\hat{\Psi}_ {\uparrow}({\bf
r}_{2})\right \rangle$, above $T_c$, such as of Ref.
\cite{Kiv,wang1}, is difficult to reconcile
 with the extremely sharp resistive and magnetic transitions at $T_{c}$ in single crystals of cuprates.
 Above $T_c$ the uniform magnetic susceptibility is paramagnetic and the resistivity is perfectly 'normal',
  showing only a few percent positive or negative magnetoresistivity. Both in-plane \cite{mac,boz,fra}
  and out-of-plane \cite {alezav} resistive transitions remain sharp in the magnetic field in
  high quality samples providing a reliable determination of a genuine $H_{c2}(T)$.

While a significant fraction of research in the field suggests
that the superconducting transition is only a phase ordering so
that a superconducting order parameter $F({\bf r}_{1},{\bf
r}_{2})$ remains nonzero above (resistive) $T_c$, we have recently
explained   the unusual Nernst signal (which is one of the key
experiments supporting that viewpoint)  as the normal state
phenomenon \cite{NERNST}.

Cuprates are known to be non-stoichiometric compounds. Moreover,
 undoped cuprates are insulators and their superconductivity
appears as a result of doping, which inevitably introduces
additional disorder. Because of these reasons, the theory of doped
semiconductors  might provide an adequate description of the
normal state kinetic properties of cuprates. Carriers in doped
semiconductors and disordered metals occupy states localised by
disorder and itinerant Bloch-like states. Both types of carriers
contribute to the transport properties, if the chemical potential
$\mu$ (or the Fermi level) is close to the energy, where the
lowest itinerant state appears (i.e. to the mobility edge).
Superconducting cuprates are among such poor conductors.

When the chemical potential is near the mobility edge, and  the
effective mass approximation is applied, there is no Nernst signal
from itinerant carriers alone, because of a so-called Sondheimer
cancellation  \cite{sond}. However, when the localised carriers
contribute to the longitudinal transport, a finite positive Nernst
signal $ e_y \equiv -E_y/\triangledown_x T$ appears as
\cite{NERNST}
\begin{equation}
{e_y\over{\rho}}={\frac{k_{B}}{{e}}}r\theta \sigma _{l},
\end{equation}
where $\rho=1/[(2s+1)\sigma_{xx}]$ is the resistivity, $s$ is the
carrier spin, $r$ is about a constant ($r$$\approx$14.3 for
fermions $s$=1/2, $r$$\approx$2.4 for bosons $s$=0), and $\Theta$
is the Hall angle. Here $\sigma_{xx}$ is the conductivity of
itinerant carriers above the mobility edge, and $\sigma_l$ is the
conductivity of localised carriers below the edge, which obeys the
Mott's law, $\sigma_l= \sigma_0 \exp \left[-(T_0/T)^{1/3}\right]$
($\sigma_0$ is about a constant).  In two dimensions  $T_0 \approx
8 \alpha^2/(k_BN_l)$, where $N_l$ is the DOS at the Fermi
level\cite{mot,shk,tok}.

In sufficiently strong magnetic field the radius of a shallow
'impurity' state $\alpha^{-1}$ is about the magnetic length,
$\alpha \approx (eB)^{1/2}$.  Then the normal state Nernst signal
is given by
\begin{equation}
{e_y\over{B\rho}}= a(T) \exp\left[-b(B/T)^{1/3}\right],
\end{equation}
where $a(T) \propto T^{-6}$ \cite{NERNST} and
$b=2[e/(k_BN_l)]^{1/3}$ is a constant.  As  shown in Ref.
\cite{NERNST}  a single-parameter Eq.(6) is in excellent
quantitative agreement with the experimental data \cite{xu,wang1}
above the resistive critical temperature $T_c(B)$, Fig.4.

As far as statistics of carriers is concerned, the bipolaron or
'preformed boson' picture is strongly supported by many other data
in underdoped and optimally doped cuprates \cite{alebook}, while
overdoped cuprates might be on the Fermi-liquid side. Unlike any
fluctuating preformed pair scenario, eg. \cite{levin}, bosons in
our model are perfectly stable, and there is no off-diagonal
order  above their Bose condensation temperature, $T_c$,
 There is  no off-diagonal order parameter above $T_c$ in the
overdoped Fermi-liquid either. In both cases the  localization of
carriers by  disorder is essential. It is responsible for the
strong Nernst signal dependence on the magnetic field, Fig.4. It
can be also responsible  for a low value of the product of the
thermopower and the Hall angle $S \tan \Theta$ in some underdoped
samples, where the contribution to the thermopower from itinerant
carriers  can be almost cancelled by the opposite sign contribution
from the localised carriers, so that $S \propto T$ at low temperatures. For two-dimensional bosons it is feasible  even if $\sigma_{xx}\gg \sigma_{l}$\cite{NERNST}.
When it happens, the Nernst signal is given by $e_y=\rho
\alpha_{xy}$, where $\alpha_{xy} \propto \tau^2$. Different from
that of fermions, the relaxation time of bosons is enhanced
critically near the Bose-Einstein condensation temperature,
$T_c(B)$, $\tau\propto[T-T_c(B)]^{-1/2}$, as in atomic Bose-gases
\cite{bec}. Providing $S\tan\Theta_H \ll e_y$, this critical
enhancement of the relaxation time describes well the temperature
dependence of $e_y$ in a few Bi2201 close to $T_c(B)$. 

  Superconductivity can be readily suppressed by the magnetic field in heavily underdoped
cuprates to
 reveal a true normal state down to zero temperature. The normal state in-plane resistivity $\rho(T)$ shows an insulating behaviour at low temperatures, which has been understood as the result of  elastic (impurity) scattering of nondegenerate carriers above the mobility edge \cite{alelog}. Combining $\rho(T) \propto 1/\tau$, $S\tan \Theta \propto T \tau$, and $e_y+S \tan \Theta \propto \tau$ one obtains  
\begin{equation}
 S\tan \Theta \propto {T\over{\rho(T)}},
 \end{equation}
 and   
 \begin{equation}
e_y \propto {T_0-T \over{\rho(T)}},
 \end{equation}
where $T_0$ is a fitting parameter. I believe that Eq.(7) and Eq.(8) can account for the experimental Nernst and $S\tan \Theta$ data in  underdoped La$_{1.92}$Sr$_{0.08}$CuO$_{4}$ \cite{cap} and  La$_{1.94}$Sr$_{0.06}$CuO$_{4}$ \cite{capcom} with  the field-induced insulating behaviour of $\rho(T)$. The unusual large Nernst signal appears due to the absence of the electron-hole symmetry near the mobility edge in the transverse transport. Our results strongly support any
microscopic theory of cuprates, which describes the state above
the resistive and magnetic phase transition as perfectly 'normal',
$F(\mathbf{r,r^{\prime }})$=$0$. Unlike \cite{wang1}, our model
does not require a radical revision of the magnetic phase diagram
of cuprates \cite{zavkabale}. Also an unusual diamagnetic response of underdoped cuprates experimentally measured somewhat above $T_c(B)$ \cite{ott,jap,ando}  can be understood as the Landau diamagnetism  of \emph{normal} state bosons \cite{dia}.

 \section{Checkerboard modulations of tunnelling DOS}

 There are more complicated
  deviations from the conventional Fermi/BCS-liquid behaviour than the normal state
  pseudogaps. Recent studies of the gap function revealed  two distinctly
 different gaps with different magnetic field and temperature
 dependence   \cite{deu,kras,kab2}, and
 the checkerboard spatial modulations of the tunnelling DOS,
  with \cite{hoff} and without \cite{hoff2,kapit} applied
magnetic fields.

We have proposed a simple phenomenological model \cite{aleand}
explaining two different gaps in  cuprates. The main assumption,
supported by a parameter-free estimate  of the Fermi energy
\cite{alefermi}, is
 that the attractive potential is large compared with the renormalised Fermi
 energy, so that the ground state is the Bose-Einstein condensate of
 tightly bound real-space pairs. Here I calculate the single particle DOS
 of  strong-coupling (bosonic) superconductors
 by solving the inhomogeneous Bogoliubov-de Genes (BdG) equations \cite{alecon}.

The anomalous Bogoliubov-Gor'kov average $F({\bf r}_{1},{\bf
r}_{2})$ depends on the relative coordinate ${\bf \rho =r}_{1}-{\bf r}%
_{2} $ of two electrons (holes), described by field operators
$\hat{\Psi} _{s}({\bf r})$,
 and on the center-of-mass coordinate $%
{\bf R}=({\bf r}_{1}+{\bf r}_{2})/2$. Its Fourier transform, $f({\bf %
k,K})$, depends on the relative momentum ${\bf k}$ and on the
center-of-mass momentum ${\bf K.}$ In the BCS theory ${\bf K=}0$,
and the Fourier transform of the order parameter is
proportional to the gap in the quasi-particle excitation spectrum, $f({\bf k,K%
})\varpropto \Delta _{{\bf k}}$. Hence the symmetry of the order
parameter and the symmetry of the gap are the same in the
weak-coupling regime. Under the rotation of the coordinate system, $\Delta _{%
{\bf k}}$ changes its sign, if the Cooper pairing appears in the
d-channel.

On the other hand,  the symmetry of the order parameter could be
different from the `internal' symmetry of the pair wave-function,
and from the symmetry of a single-particle excitation gap in the
strong-coupling regime \cite{alebook}. Real-space pairs might have
an unconventional symmetry due to a specific symmetry of the
pairing potential as in the case of the Cooper pairs, but in any
case the ground state and DOS are homogeneous, if pairs are
condensed with ${\bf K}=0$. However, if a pair band dispersion has
its minima at finite ${\bf K}$ in the center-of-mass Brillouin
zone, the Bose condensate is inhomogeneous. In particular,   the
center-of-mass bipolaron energy bands could have their minima at
the Brillouin zone boundaries at ${\bf K}=(\pi,0)$ and  three
other equivalent momenta \cite{alesym} (here and further I take
the lattice constant $a=1$). These four states are degenerate, so
that
the condensate wave function $\psi({\bf m})$ in the real (Wannier) space, ${\bf m}%
=(m_{x},m_{y}),$ is their superposition,
\begin{equation}
\psi({\bf m})=\sum_{{\bf K}=(\pm \pi ,0),(0,\pm \pi )}b_{{\bf K}%
}e^{-i{\bf K\cdot m}},
\end{equation}
where $b_{{\bf K}}=\pm \sqrt{n_{c}}/2$ are $c$-numbers,  and
$n_c(T)$ is the  atomic density of the Bose-condensate. The
superposition, Eq.(9), respects the time-reversal and parity
symmetries, if
\begin{equation}
\psi ({\bf m})=\sqrt{n_{c}}\left[ \cos (\pi m_{x})\pm \cos (\pi
m_{y})\right] .
\end{equation}
The order parameter, Eq.(10), has $d$-wave symmetry  changing sign
in the real space, when the lattice is rotated by $\pi /2$. This
symmetry is
entirely due to the pair-band energy dispersion with four minima at ${\bf %
K} \neq 0$, rather than due a  specific pairing potential. It
reveals itself as a {\it checkerboard} modulation of the hole
density with two-dimensional patterns, oriented along the
diagonals.  From this insight  one can expect  a fundamental
connection between stripes detected by different techniques
\cite{bia}  and the symmetry of the order parameter in cuprates
\cite{alesym}.

Now  let us take into account that in the superconducting state
($T<T_c$)  single-particle excitations
 interact with the pair condensate via the same  attractive potential, which forms the pairs \cite{aleand}.
 The Hamiltonian  describing
the interaction of  excitations with the pair Bose-condensate in
the Wannier representation is
\begin{equation}
 H = -\sum_{s,{\bf m,n}}[t({\bf m-n})+\mu \delta_{\bf m,n}]c^{\dagger}_{s \bf m}c_{s \bf
 n}
 + \sum_{\bf m}[\Delta({\bf m})c^{\dagger}_{\uparrow \bf m}c_{\downarrow \bf
 m}+H.c.],
 \end{equation}
 where $s= \uparrow,\downarrow$ is the
 spin,   and $\Delta({\bf m}) \propto \psi ({\bf
 m})$. Applying  equations of motion for  the Heisenberg operators
 $\tilde{c}^{\dagger}_{s \bf m}(t)$ and $\tilde{c}_{s \bf
 m}(t)$, and the Bogoliubov transformation \cite{bog}
 \begin{equation}
\tilde{c}_{\uparrow \bf m}(t)=\sum_{\nu} [u_{\nu}({\bf m})
\alpha_{\nu} e^{-i\epsilon_{\nu}t} +v_{\nu}^* ({\bf
m})\beta_{\nu}^{\dagger} e^{i\epsilon_{\nu}t}],
\end{equation}
\begin{equation}
\tilde{c}_{\downarrow \bf m}(t)=\sum_{\nu} [u_{\nu}({\bf m})
\beta_{\nu} e^{-i\epsilon_{\nu}t} - v_{\nu}^* ({\bf
m})\alpha_{\nu}^{\dagger} e^{i\epsilon_{\nu}t}],
\end{equation}
one  obtains BdG equations describing  the single-particle
excitation spectrum,
\begin{equation}
\epsilon u({\bf m})=-\sum_{\bf n} [t({\bf m-n})+\mu \delta_{\bf
m,n}]u({\bf n}) + \Delta({\bf m})v ({\bf m}),
\end{equation}
\begin{equation}
-\epsilon v({\bf m})=-\sum_{\bf n} [t({\bf m-n})+\mu \delta_{\bf
m,n}]v({\bf n}) + \Delta({\bf m})u({\bf m}),
\end{equation}
where  excitation quantum numbers $\nu$ are omitted for
transparency. Different from the conventional BdG equations in the
weak-coupling limit, there is virtually no feedback  of  single
particle excitations on the off-diagonal potential, $\Delta({\bf
m})$, in the strong-coupling regime. The number of these
excitations is low at temperatures below $\Delta_p/k_B$, so that
the coherent potential $\Delta ({\bf m})$ is an external (rather
than a self-consistent) field,  solely determined by the pair Bose
condensate \cite{aleand}.

While the analytical solution is not possible for any arbitrary
off-diagonal interaction $\Delta({\bf m})$, one can readily solve
the infinite system of discrete equations (14,15) for a periodic
$\Delta({\bf m})$ with a period commensurate with the lattice
constant. For example,
\begin{equation}
\Delta({\bf m})= \Delta_c [e^{i\pi m_x} - e^{i\pi m_y}],
\end{equation}
 corresponds to the pair condensate at ${\bf K}=(\pm\pi,0)$
and $(0,\pm\pi)$, Eq.(10), with a temperature dependent (coherent)
$\Delta_c \propto \sqrt {n_c(T)}$. In this case the quasi-momentum
${\bf k}$ is the proper quantum number, $\nu= {\bf k}$, and the
excitation wave-function is a superposition of plane waves,
\begin{eqnarray}
u_{\nu}({\bf m}) &=&u_{{\bf k}}e^{i{\bf k}\cdot {\bf m}}+\tilde{u}_{{\bf k}}e^{i({\bf k-g})\cdot {\bf m}}, \\
v_{{\nu}}({\bf m}) &=&v_{{\bf k}}e^{i{({\bf k-g}_x})\cdot {\bf
m}}+\tilde{v}_{{\bf k}}e^{i({{\bf k-g}_y})\cdot {\bf m}}.
\end{eqnarray}
Here ${\bf g}_x=(\pi,0)$, ${\bf g}_y=(0,\pi)$, and ${\bf
g}=(\pi,\pi)$ are reciprocal doubled lattice vectors. Substituting
Eqs.(17) and (18) into Eqs.(14,15) one obtains four coupled
algebraic equations,
\begin{eqnarray}
\epsilon _{\bf k}u_{\bf k}&=&\xi_{\bf k}u_{\bf k}-\Delta_c (v_{\bf k}-\tilde{v}_{\bf k}), \\
\epsilon _{\bf k}\tilde{u}_{\bf k}&=&\xi_{\bf k-g}\tilde{u}_{\bf
k}+\Delta_c (v_{\bf k}-\tilde{v}_{\bf k}),
\\
-\epsilon_{\bf k}v_{\bf k}&=&\xi_{{\bf k-g}_x}v_{\bf k}+\Delta_c
(u_{\bf k}-\tilde{u}_{\bf k}), \\ -\epsilon_{\bf k}\tilde{v}_{\bf
k}&=&\xi_{{\bf k-g}_y}\tilde{v}_{\bf k}-\Delta_c (u_{\bf
k}-\tilde{u}_{\bf k}) ,
\end{eqnarray}
where $ \xi_{\bf k}=-\sum_{\bf n}t({\bf n}) e^{i{\bf k \cdot n}}
-\mu$. The determinant of the system (19-22) yields the following
equation for  the energy spectrum $\epsilon$:
\begin{eqnarray}
&&(\epsilon-\xi_{\bf k})(\epsilon-\xi_{\bf
k-g})(\epsilon+\xi_{{\bf k-g}_x})(\epsilon+\xi_{{\bf k-g}_y})\cr
&=&\Delta_c^2(2\epsilon+\xi_{{\bf k-g}_x}+\xi_{{\bf
k-g}_y})(2\epsilon-\xi_{\bf k}-\xi_{\bf k-g}).
\end{eqnarray}
Two positive  roots for $\epsilon$ describe the single-particle
excitation spectrum. Their calculation is rather cumbersome, but
not in the extreme strong-coupling limit, where the pair binding
energy $2\Delta_p$ is large compared with $\Delta_c$ and with the
single-particle bandwidth $w$. The
 chemical potential in this limit is pinned
 below a single-particle band edge, so that $\mu=-(\Delta_p+w/2)$ is negative, and
its magnitude is large compared with $\Delta_c$.
 Then the right
hand side in Eq.(23) is a perturbation, and the spectrum is
\begin{eqnarray}
\epsilon_{1\bf k}&\approx& \xi_{\bf k} -{\Delta_c^2\over{\mu}}, \\
\epsilon_{2\bf k}&\approx& \xi_{\bf k-g} -{\Delta_c^2\over{\mu}},
\end{eqnarray}

If a metallic tip is placed at the point ${\bf m}$ above  the
surface of a sample, the STM current $I(V, {\bf m)}$ creates an
electron (or hole) at this point. Applying the Fermi-Dirac golden
rule and the Bogoliubov transformation, Eq.(12,13), and assuming
that the temperature is much lower than $\Delta_p/k_B$ one readily
obtains the tunnelling conductance
\begin{equation}
\sigma (V, {\bf m})\equiv {dI(V, {\bf m)}\over {dV}}\propto
\sum_{\nu} | u_{\nu}({\bf m})|^2 \delta(eV-\epsilon_{\nu}),
 \end{equation}
 which is a local excitation DOS. The solution  Eq.(17) leads to a spatially modulated
 conductance,
\begin{equation}
\sigma(V, {\bf m)}= \sigma_{reg}(V) +\sigma_{mod} (V) \cos(\pi
m_x+\pi m_y).
\end{equation}
The smooth (regular) contribution is
\begin{equation}
\sigma_{reg}(V)=\sigma_0 \sum_{{\bf k},r=1,2}(u_{r{\bf k}}^2
+\tilde{u}_{r {\bf k}}^{2}) \delta(eV-\epsilon_{r{\bf k}}),
\end{equation}
and the amplitude of the modulated contribution is
\begin{equation}
\sigma_{mod}(V)= 2\sigma_0 \sum_{{\bf k},r=1,2}u_{r{\bf
k}}\tilde{u}_{r {\bf k}} \delta(eV-\epsilon_{r{\bf k}}),
\end{equation}
where $\sigma_0$ is a constant. Conductance modulations reveal a
checkerboard pattern, as the Bose condensate itself, Eq.(10),
\begin{equation}
{\sigma-\sigma_{reg}\over{\sigma_{reg}}}=A \cos(\pi m_x+\pi m_y),
\end{equation}
where
 \begin{eqnarray}
 &&A=2 \sum_{\bf k} \left[u_{1
\bf k}\tilde{u}_{1\bf k} \delta (eV-\epsilon_{1 \bf k})+ u_{2\bf
k}\tilde{u}_{2 \bf k}  \delta (eV-\epsilon_{2 \bf k})\right]/\cr
&& \sum_{\bf k} \left [(u_{1 \bf k}^2 +\tilde{u}_{1 \bf k}^2)
\delta (eV-\epsilon_{1 \bf k})+(\tilde{u}_{2 \bf k}^2+u_{2 \bf
k}^2) \delta (eV-\epsilon_{2 \bf k})\right] \nonumber
\end{eqnarray}
is the amplitude  of  modulations  depending on the voltage $V$
and temperature. An  analytical result can be obtained in the
strong-coupling limit with the excitation spectrum  given by Eqs.
(24,25) for the voltage  near the threshold, $eV\approx \Delta_p$.
In this case only  states near bottoms of each excitation band
contribute to the integrals in Eq.(30), so that
\begin{equation}
\tilde{u}_{1\bf k}={\xi_{\bf k}-\epsilon_{1 \bf
k}\over{\epsilon_{1 \bf k}-\xi_{\bf k-g}}}u_{1\bf k} \approx
-u_{1\bf k}{\Delta_c^2\over{ \mu w}}\ll {u}_{1\bf k},
\end{equation}
and
\begin{equation}
u_{2\bf k}={\xi_{\bf k-g}-\epsilon_{2 \bf k}\over{\epsilon_{2 \bf
k}-\xi_{\bf k}}}\tilde{u}_{2\bf k} \approx -\tilde{u}_{2\bf
k}{\Delta_c^2\over{ \mu w}}\ll \tilde{u}_{2\bf k}.
\end{equation}
Substituting  these expressions into $A$, Eq.(30), yields in the
lowest order of $\Delta_c$,
\begin{equation}
A\approx -{2\Delta_c^2\over{ \mu w}}.
\end{equation}
The result, Eq.(30), is  reminiscent of STM data
\cite{hoff,hoff2,kapit,fu}, where  spatial checkerboard
modulations of $\sigma$ were observed in a few cuprates. Both commensurate and incommensurate 
 modulations were found depending on sample composition. In our model
the period is determined by the center-of mass wave vectors ${\bf
K}$ of the Bose-condensed preformed pairs. While the general case
has to be solved numerically, the perturbation result, Eq.(30) is
qualitatively applied for any ${\bf K}$ at least close to $T_{c}$,
where the coherent gap is small, if one replaces $\cos(\pi m_x+\pi
m_y)$ by $\cos(K_{x} m_x+K_{y} m_y)$. 
Different from any other scenario, proposed so far, the hole
density, which is about twice of the condensate density at low
temperatures, is spatially modulated with the  period determined
by the inverse wave vectors corresponding to the center-of-mass
pair band-minima. This 'kinetic' interpretation of charge
modulations in cuprates was originally proposed \cite{alesym} before
STM results became available. It could account for those DOS modulations in  superconducting samples, which disappear above $T_c$ because  the coherent gap $\Delta_c(T)$ vanishes, so that
$A=0$ in Eq.(30). Indeed some 
inelastic
neutron scattering experiments show that
incommensurate inelastic peaks are observed $only$ in the $%
superconducting$ state of high-$T_c$ cuprates \cite{bou}. The vanishing at $T_{c}$ of 
incommensurate  peaks is inconsistent with any other stripe
picture, where a characteristic distance needs to be observed in
the normal state as well.  On the other hand some  STM studies (see, for example \cite{ver}) report
incommensurate and commensurate DOS modulations somewhat above $T_c$, in particular, in heavily underdoped cuprates \cite{hancond}. I believe that those modulations are due to a single-particle band structure and impurity states near the top of the valence band in doped charge-transfer insulators, rather than a signature of any cooperative phenomenon.

In conclusion, the strong-coupling Fr\"ohlich-Coulomb model, Eq.(1), 
 links charge heterogeneity, pairing, and pseudo-gaps as
manifestations of the strong electron-phonon attractive
 interaction in narrow bands of doped Mott-Hubbard insulators.

The author acknowledges support of this work  by the Leverhulme
Trust (London) Grant No. F/00261/H ,  The Royal Society, and by EPSRC (UK) Grant No.  EP/C518365/1.

\newpage

\textbf{Figure captions:}

\vspace{1cm}
\textbf{Fig.1.} Top panels: small Holstein polaron band dispersions along
the main
directions of the two-dimensional Brillouin zone. Left: $\protect\lambda %
=1.1 $, right: $\protect\lambda =1.2$. Filled symbols is $\protect\omega %
=0.70\,t$, open symbols is $\protect\omega =0.66\,t$. Lower
panels: the band
structure isotope exponent for $\protect\lambda =1.1$ (left) and $\protect%
\lambda =1.2$ (right).

\textbf{Fig.2.} Top panels: small Fr\"ohlich polaron band dispersions
along the
main directions of the two-dimensional Brillouin zone. Left: $\protect%
\lambda = 2.5$, right: $\protect\lambda = 3.0$. Filled symbols is $\protect%
\omega = 0.70 \, t$, open symbols is $\protect\omega = 0.66 \, t$.
Lower panels: the band structure isotope exponent for
$\protect\lambda = 2.5$ (left) and $\protect\lambda = 3.0$
(right).

\textbf{Fig.3.} The Hall Lorenz number $L_H$ \cite{lor} fits the
experiment (YBa$_2$Cu$_3$O$_{6.95}$ \cite{zha}). Charge and spin
pseudogaps are taken as 675 K and 150 K, respectively, and the
ratio of the polaron and bipolaron Hall angles is 0.36. The inset
gives the ratio of Hall Lorenz number to Lorenz number in the
model.

\textbf{Fig.4.} Eq.(6) fits the experimental signal
(symbols) in overdoped La$_{1.8}$Sr$_{0.2}$CuO$_{4}$ \cite{wang1} with
$b=7.32$(K/Tesla)$^{1/3}$. Inset shows $a(T)$ obtained from the
fit (dots) together with $a\propto T^{-6}$ (line).

\newpage

\begin{figure}[tbsp]
\begin{center}
\includegraphics[angle=-00,width=0.40\textwidth]{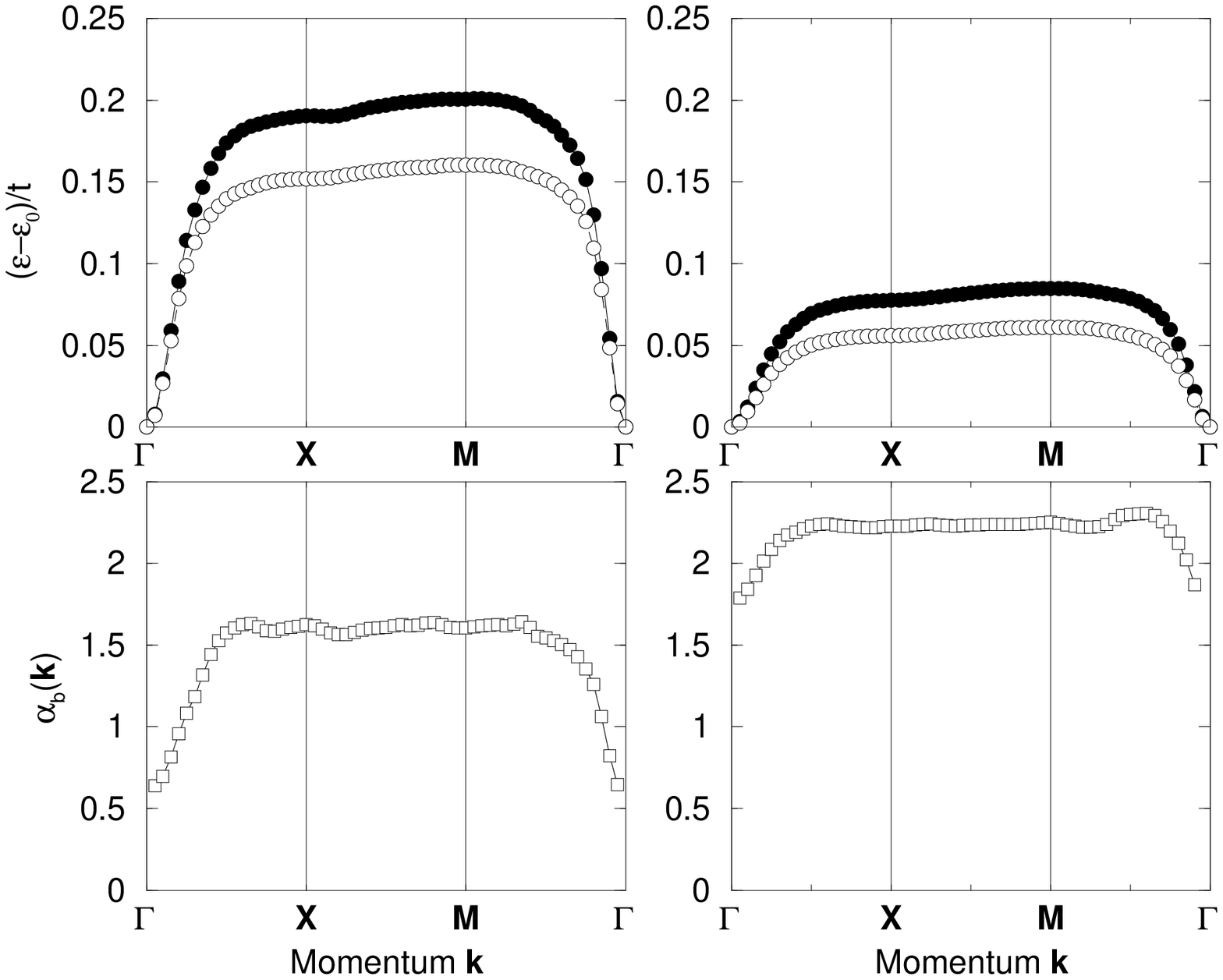}
\caption{}
\end{center}

\end{figure}

\begin{figure}[tbsp]
\begin{center}
\includegraphics[angle=-00,width=0.40\textwidth]{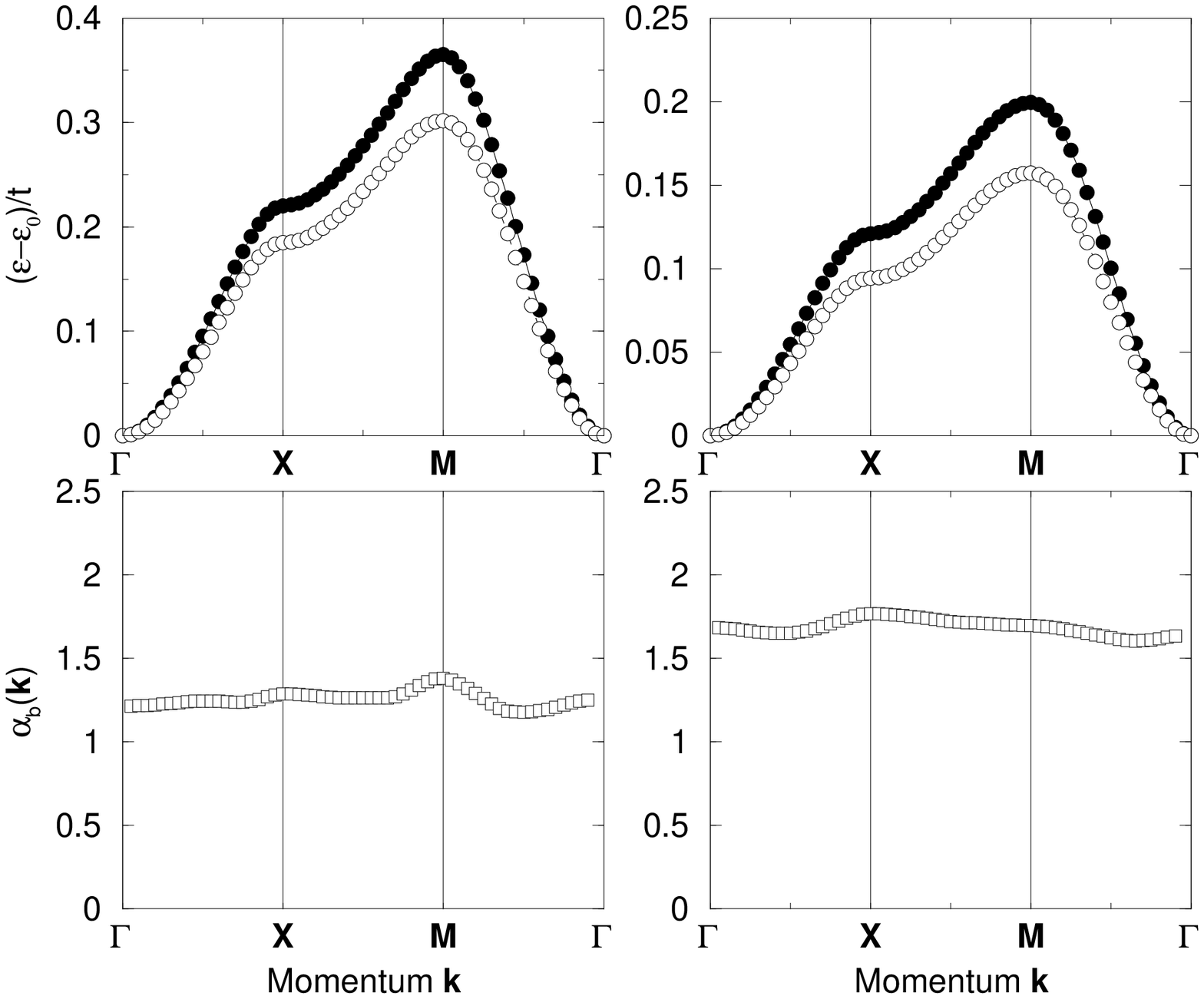}
\caption{}
\end{center}

\end{figure}

\begin{figure}[tbsp]
\begin{center}
\includegraphics[angle=-00,width=0.50\textwidth]{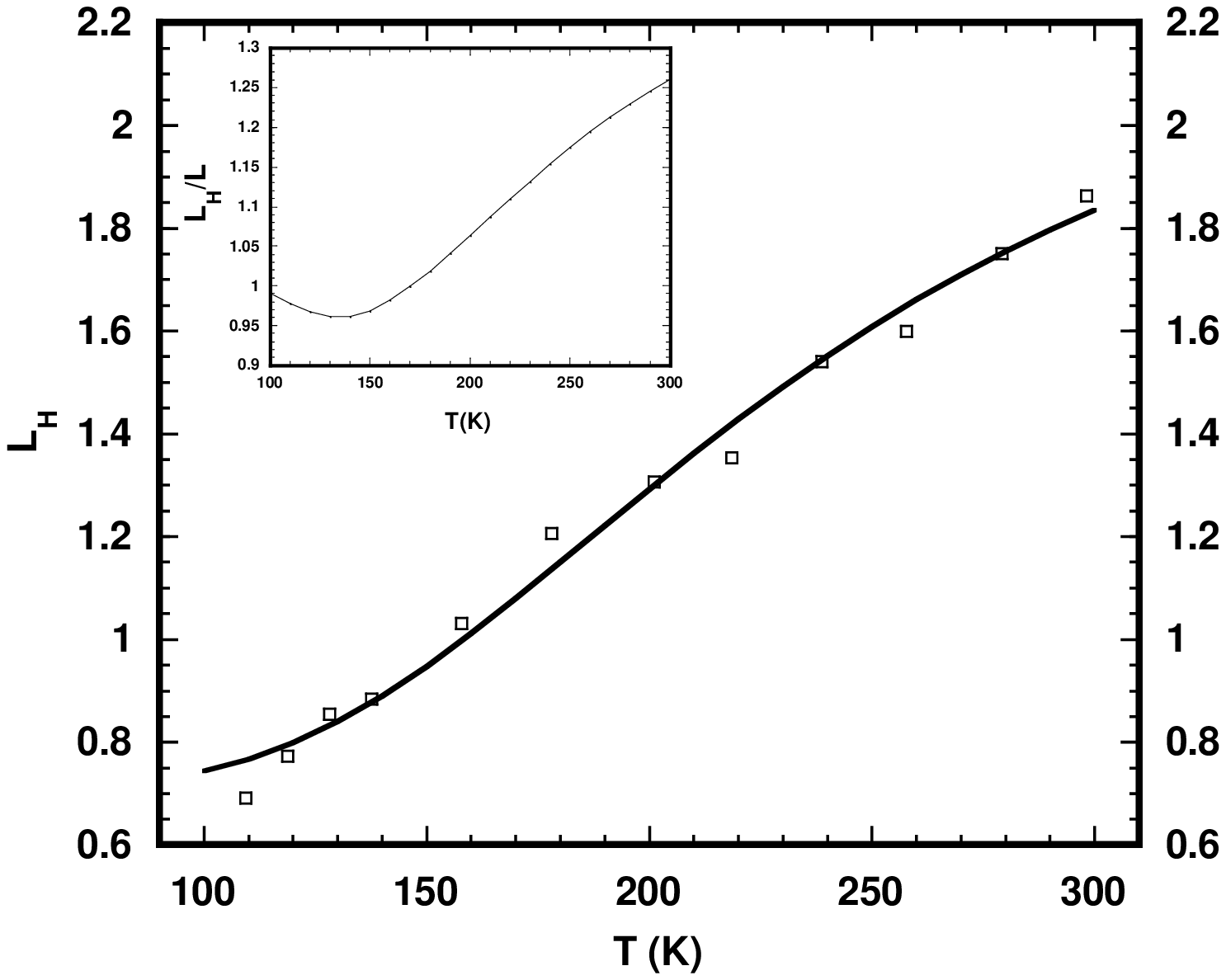}
\caption{}
\end{center}
\end{figure}
\begin{figure}
\begin{center}
\includegraphics[angle=-0,width=0.47\textwidth]{Fig.4.EPS}
\caption{}
\end{center}
\end{figure}


\begin{thebibliography}{99}
\bibitem{and2} P.W. Anderson, \emph{The Theory of Superconductivity in the High T$_c$ Cuprates}
(Princeton University Press, Princeton, 1997).
\bibitem{kiv}
E.W. Carlson, V.J. Emery, S.A. Kivelson, and D. Orgad, in
\emph{The Physics of Superconductors}, Vol.2, eds. K.H. Bennemann
and J.B. Ketterson (Springer-Verlag, NY, 2004).
\bibitem{koh}
W. Kohn and J.M. Luttinger, Phys. Rev. Lett. {\bf 15}, 524 (1965).
\bibitem{gab}
A.M. Gabovich, A.I. Voitenko, J.F. Annett, and M. Ausloos, 2001
 Supercond. Sci. Technol. {\bf 14}, R1 (2001).
\bibitem{mott}
N.F. Mott, {\it Metal-Insulator Transitions, 2nd ed}. (
Taylor\&Francis, London, 1990).
\bibitem{scher}
A. Scherman  and M. Scrieber, Phys. Rev. B{\bf 52}, 10621 (1995).
\bibitem{lau}
R.B. Laughlin,  cond-mat/0209269.
\bibitem{bas}
J.H. Kim, B.J. Feenstra, H.S. Somal, D. van der Marel, W.Y. Lee,
A.M. Gerrits, and A. Wittlin, Phys. Rev. B{\bf 49}, 13065 (1994).
\bibitem{alebook} A.S. Alexandrov, \emph{Theory of Superconductivity: from Weak to
Strong Coupling} (IoP Publishing, Bristol-Philadelphia, 2003).
\bibitem{alekor}
A.S. Alexandrov and P.E. Kornilovitch,  J. Phys.: Condens. Matter
\textbf{14}, 5337 (2002).
\bibitem{iso} E. Maxwell, Phys. Rev. \textbf{78}, 477
(1950); C. A. Reynolds, B. Serin, W. H. Wright, and L. B. Nesbitt,
ibid 487 (1950).
\bibitem{aleiso}  A.S. Alexandrov, Phys. Rev. B\textbf{46},
14932 (1992).
\bibitem{guo0}
G. Zhao and D. E. Morris,
Phys. Rev. B {\bf 51}, 16487 (1995).
\bibitem{guo}  G. Zhao, M.B. Hunt, H. Keller, and K.A. M\"uller, Nature $\mathbf{385}$, 236 (1997).
\bibitem{kha}  R. Khasanov, D.G. Eshchenko, H. Luetkens, E. Morenzoni,
T. Prokscha, A. Suter, N. Garifianov, M. Mali, J. Roos, K. Conder,
and H. Keller, Phys. Rev. Lett. \textbf{92}, 057602 (2004).
\bibitem{mig}  A.B. Migdal, Zh. Eksp. Teor. Fiz. \textbf{34}, 1438 (1958)
(Sov. Phys. JETP \textbf{7}, 996 (1958)).
\bibitem{lan0}
A. Lanzara, P.V. Bogdanov, X.J. Zhou, S.A. Kellar, D.L. Feng, E.D.
Lu, T. Yoshida, H. Eisaki, A. Fujimori, K. Kishio, J.I. Shimoyana,
T. Noda, S. Uchida, Z. Hussain, and Z.X. Shen, Nature {\bf 412},
510 (2001).
\bibitem{shencon}
X.J. Zhou, J. Shi, T. Yoshida, T. Cuk, W. L. Yang, V. Brouet, J.
Nakamura, N. Mannella, S. Komiya, Y. Ando, F. Zhou, W. X. Ti, J.
W. Xiong, Z. X. Zhao, T. Sasagawa, T. Kakeshita, H. Eisaki, S.
Uchida, A. Fujimori, Zhenyu Zhang, E. W. Plummer, R. B. Laughlin,
Z. Hussain, and Z.-X. Shen, cond-mat/0405130.
\bibitem{lan}
G-H. Gweon, T. Sasagawa, S.Y. Zhou, J. Craf, H. Takagi, D.-H. Lee,
and A. Lanzara, Nature, {\bf 430}, 187 (2004).
\bibitem{alezhao}  A. S. Alexandrov, G.-m. Zhao, H. Keller, B. Lorenz, Y. S. Wang, and C. W. Chu, Phys. Rev. B \textbf{64}, 140404(R) (2001).
\bibitem{alekor1}  A.S. Alexandrov and P.E. Kornilovitch, Phys. Rev. Lett.
\textbf{82}, 807 (1999).
\bibitem{alekor2}  A.S. Alexandrov and P.E. Kornilovitch,
Phys. Rev. B{\bf 70}, 224511 (2004).
\bibitem{kor2}  P.E. Kornilovitch, EuroPhys. Lett. \textbf{59}, 735 (2002).
\bibitem{dev}  J. Tempere and J.T. Devreese, Phys. Rev. B \textbf{64},
104504 (2001).
\bibitem{aledent}  A.S. Alexandrov and C.J. Dent, Phys. Rev. B \textbf{60},
15414 (1999).
\bibitem{tim}  T. Timusk and B. Statt, Rep. Prog. Phys. {\bf 62}, 61 (1999).
\bibitem{NEV}  A.S. Alexandrov and N.F. Mott, Phys. Rev. Lett, {\bf 71},
1075 (1993).
\bibitem{zha}  Y. Zhang, N.P. Ong, Z.A. Xu, K. Krishana, R. Gagnon, and L.
Taillefer, Phys. Rev. Lett., {\bf 84}, 2219 (2000).
\bibitem{lor} K.K. Lee, A.S. Alexandrov and W.Y. Liang, Phys. Rev.
Lett. {\bf 90}, 217001 (2003); ibid, Eur. Phys. J. B{\bf 39}, 459
(2004).
\bibitem{xu}  Y.\,Wang, S.\,Ono, Y.\,Onose, G.\,Gu, Y.\,Ando, Y.\,Tokura,
S.\,Uchida, N.P.\,Ong, Science \textbf{299}, 86 (2003).
\bibitem{wang1}  Y.\,Wang, N.P.\,Ong, Z.A.\,Xu, T.\,Kakeshita, S.\,Uchida, D.A.\,Bonn,
R.\,Liang, W.N.\,Hardy, Phys.\,Rev.\,Lett. \textbf{88}, 257003 (2002).
\bibitem{Kiv}  V.J. Emery, S.A. Kivelson, Nature, \textbf{374}, 434
(1995).
\bibitem{mac}  A.P.\,Mackenzie, S.R.\,Julian, G.G.\,Lonzarich,
 A.\,Carring\-ton, S.D.\,Hughes, R.S.\,Liu, D.C.\,Sinclair,
Phys.\,Rev.\,Lett. \textbf{71}, 1238 (1993)
\bibitem{boz}  M.S. Osofsky, R. J. Soulen, S.A. Wolf, J.M. Broto, H. Rakoto,
J.C. Ousset, G. Coffe, S. Askenazy, P. Pari, I. Bozovic, J.N.
Eckstein, G. F. Virshup, Phys. Rev. Lett. \textbf{71}, 2315
(1993); ibid \textbf{72}, 3292 (1994).
\bibitem{fra}  D.D.\,Lawrie,\,J.P.\,Franck,\,J.R.\,Beamish,\,E.B.\,Molz, W.M.\,Chen,
M.J.\,Graft, J.\,Low\,Temp.\,Phys.\,\textbf{107},\,491\,(1997).
\bibitem{alezav} A.S. Alexandrov, V.N. Zavaritsky, W.Y.
Liang,  P.L. Nevsky, Phys. Rev. Lett. \textbf{76} 983 (1996).
\bibitem{NERNST}
 A. S. Alexandrov and V. N. Zavaritsky
Phys. Rev. Lett. {\bf 93} 217002 (2004).
\bibitem{sond}  E.H. Sondheimer, Proc. Roy. Soc. %(London)
\textbf{193}, 484 (1948).
\bibitem{mot} N.F. Mott, and E.A. Davis, \emph{Electronic Processes
in Non-Crystalline Materials} (Clarendon Press,Oxford, 1979).
\bibitem{shk} B.I. Shklovskii, Soviet Phys. JETP Lett. {\bf 36},
51 (1982).
\bibitem{tok} R. Mansfield and
H. Tokumoto, Phil Mag. B{\bf 48}, L1 (1983).
\bibitem{zavkabale} A.S. Alexandrov, Phys. Rev. B {\bf 48}, 10571
(1993); V.N. Zavaritsky, V.V. Kabanov, and A.S. Alexandrov, Europhys.
Lett. ${\bf 60}$, 127, (2002).
\bibitem{levin} S. Tan and K. Levin, Phys.Rev. B 69, 064510 (2004).
\bibitem{bec} T.\,Lopez-Arias and A.\,Smerzi,  Phys. Rev. A {\bf 58}, 526 (1998).
\bibitem{alelog}
A.S. Alexandrov, Phys. Lett. A{\bf 236}, 132 (1997). 
\bibitem{cap} C. Capan, K. Behnia, J. Hinderer, A.G.M. Jansen, W. Lang, C. Marcenat, C. Marin, and J. Flouquet, Phys. Rev. Lett. {\bf 88}, 056601 (2002).
\bibitem{capcom} 
C. Capan and K. Behnia, cond-mat/0501288.
\bibitem{ott}
R. Jin, A. Schilling, and H. R. Ott, Phys. Rev. B 49, 9218 (1994).
\bibitem{jap}
A. Sugimoto, I. Iguchi, T. Miyake,  and H. Sato,
Japanise Journal of Applied Physics, 
{\bf 41},  L497 (2002).
\bibitem{ando}
Y. Wang, Lu Li, M. J. Naughton, G. D. Gu, S. Uchida, and N. P. Ong, cond-mat/0503190.
\bibitem{dia}
C.J. Dent, A.S. Alexandrov and V.V. Kabanov Physica C{\bf 341}, 153 (2000).
\bibitem{deu} G. Deutscher, Nature {\bf 397}, 410 (1999).
\bibitem{kras}
V. M. Krasnov, A. Yurgens, D. Winkler, P. Delsing, and T. Claeson,
Phys. Rev. Lett. {\bf 84}, 5860 (2000).
\bibitem{kab2} J. Demsar, R. Hudej, J. Karpinski, V. V. Kabanov, and D. Mihailovic, Phys. Rev. B {\bf 63}, 054519 (2001).
\bibitem{hoff}  J. E. Hoffman, E. W. Hudson, K. M. Lang, V. Madhavan, H. Eisaki, S. Uchida, and J. C. Davis,
  Science {\bf 295}, 466 (2002).
\bibitem{hoff2} J. E. Hoffman, K. McElroy, D.-H. Lee, K. M Lang, H.
Eisaki, S. Uchida, and J. C. Davis, Science {\bf 297}, 1148
(2002).
\bibitem{kapit}  C. Howald, H. Eisaki, N. Kaneko, M. Greven, and A. Kapitulnik, Phys. Rev. B{\bf 67},
014533 (2003).
\bibitem{aleand}  A.S. Alexandrov  and A.F. Andreev, Europhys.
Lett. {\bf 54}, 373 (2001).
\bibitem{alefermi}
A.S. Alexandrov, Physica C {\bf 363}, 231 (2001).
\bibitem{alecon}
A.S. Alexandrov, cond-mat/0407401.
\bibitem{alesym}
A.S. Alexandrov, Physica C {\bf 305}, 46 (1998).
\bibitem{bia}  A. Bianconi,  J. Phys. IV France {\bf 9},
325 (1999), and references therein.
\bibitem{bog}
N. Bogoliubov, J.Phys. USSR {\bf 11}, 23 (1947).
\bibitem{fu} for an experimental summary see H.C. Fu, J.C. Davis
and D.-H Lee, cond-mat/0403001.

\bibitem{bou}  P. Bourges, Y. Sidis, H. F. Fong, L. P. Regnault, J. Bossy, A. Ivanov, and B. Keimer,
Science, {\bf 288} 1234 (2000); J. Supercond. {\bf 13}, 735
(2000).
\bibitem{ver} M. Vershinin, S. Misra, S. Ono, Y. Abe, Y. Ando, and A. Yazdani, Science {\bf 303}, 1995
(2004).
\bibitem{hancond}
T. Hanaguri, C. Lupien, Y. Kohsaka, D. -H. Lee, M. Azuma, M. Takano, H. Takagi, and J. C. Davis, cond-mat/0409102.
\end{thebibliography}
\end{document}